\documentclass[aps,pra,epsfigure,twocolumn,showpacs]{revtex4}
\usepackage{dcolumn}
\usepackage{bm}
\usepackage{graphicx}
\usepackage{amsmath}
\usepackage{latexsym}
\usepackage{amsfonts}
\usepackage{amssymb}
\usepackage{array}
\usepackage{epsfig}

\newcommand{\ket}[1]{\left\vert#1\right\rangle}

\newcommand{\bra}[1]{\left\langle#1\right\vert}

\newcommand{\nbar}{\overline{n}}
\newcommand{\expect}[3]{\langle#1|#2|#3\rangle}
\begin{document}


\title{Entanglement generation and protection by detuning modulation}
\author{M. Paternostro$^1$, M. S. Tame$^1$, G. M. Palma$^2$, and M. S. Kim$^1$}
\affiliation{$^1$School of Mathematics and Physics, Queen's University, Belfast BT7 1NN, United Kingdom\\
$^2$NEST-INFM\&{D}ipartimento di Scienze Fisiche ed Astronomiche, Universita' degli studi di Palermo, via Archirafi 36, 90123, Italy}

\date{\today}

\begin{abstract}
We introduce a protocol for steady-state entanglement generation and protection based on detuning modulation in the dissipative interaction between a two-qubit system and a bosonic mode.
The protocol is a global-addressing scheme which only requires control over the system as a whole. We describe a postselection procedure to project the register state onto a subspace of maximally entangled states. We also outline how our proposal can be implemented in a circuit-quantum electrodynamics setup.
\end{abstract}

\maketitle

\section{Introduction}


The controllable generation of entangled states has triggered a considerable amount of interest in the physics community. 
In particular, within cavity-quantum electrodynamics (cavity-QED) contexts, the achievement of two-qubit entanglement has seen a flourishing of proposals. It has been suggested to set entanglement between two remote atomic qubits by using the cancellation of which-path information relative to spontaneously emitted photons~\cite{whichpath}. The resonant interaction of a cavity field with two qubits has revealed a striking {entangling power} even when the field is prepared in an incoherent state~\cite{myungpeter}. In a similar setup, entanglement can be created through the continuous detection of the field leaking from the cavity containing the atoms~\cite{pleniohuelga}. However, qubit entanglement can also be set in a regime of dissipative dynamics, where the system at hand interacts with a structured environment~\cite{massimo} and the conditions for entanglement generation between subsystems undergoing purely dissipative dynamics have been studied~\cite{benatti,tanasficek}. Strategies for the effective engineering and simulation of such environments have subsequently been envisaged~\cite{engineered}. Most recently, schemes have been proposed for efficiently inducing discrete-variable entanglement in a bipartite system by {\it transferring} the correlation properties of a continuous-variable state~\cite{ET}. 

It is easy to recognize the importance of protocols that are able to reliably protect correlations, once they are established, from the unavoidable spoiling effects of decoherence and decay. This has led to proposals of {\it passive} as well as {\it active} schemes~\cite{misto}. The use of decoherence-free subspaces is the prototypical example of passive strategies, where the correlations set in a system can be protected by choosing a proper encoding of the information~\cite{DFS,beige}. More recently, it has been suggested to use macroscopic quantum jumps as a tool to create an entangled state of two qubits, preparing it in a dark state~\cite{metz}. One could also consider generalized dynamical-decoupling ({bang-bang}) schemes~\cite{bangbang} to actively cancel the effect of the environment on the system of interest. These strategies are appealing and intriguing from a theoretical point of view and proof-of-principle experiments have been performed, for example in a solid-state setup~\cite{solidoDFS}. However, they are still far from complete and too demanding for the current state of the art. 

In this paper we find a strategy, based on dissipative qubit-bus dynamics, enabling the simultaneous {generation} of two-qubit entanglement and {protection} against both qubit and bus losses. No structured-bath engineering is required in our scheme. The protection is achieved by using a simple global addressing of the register, a feature which relaxes the usually assumed requirement of single-qubit addressing at the center of many dynamical-decoupling protocols and brings our scheme closer to experimental feasibility. While a quantitative analysis is deferred until later, here we briefly provide an intuitive picture of the mechanism behind our proposal. 
It is known from the study of the so-called Dicke model~\cite{dicke} that a bipartite qubit system, prepared with the qubits in their excited state and exposed to the fluctuations of a common reservoir, soon decays via the channel given by the symmetric state $\ket{s}=(1/\sqrt 2)(\ket{01}+\ket{10})$ (with $\ket{0}$ and $\ket{1}$ the single-qubit logical levels) into the total ground state of the qubits. Thus, there is a transient period in the dynamics of the two qubits when a maximally entangled component is involved in the state of the system. However, the system does not exhibit entanglement because of a competition between the fading symmetric state and the increasingly populated collective ground state. In order to reverse this situation, the influence of $\ket{s}$ has to be emphasized and {stabilized} with time. The protection from environmental effects is then achieved by this stabilization and also by simultaneously inducing a relative phase between the qubits. In proper conditions, this results in part of the population of the symmetric state being moved into the antisymmetric state $\ket{a}=({1/\sqrt 2})(\ket{01}-\ket{10})$, which is decoupled from the decay mechanism (it is a {subradiant} state)~\cite{dicke}. In this paper we show that an entangled steady state is produced by modulating a detuning between the common bus and the register. The results depend on the temporal profile of the modulation, which represents a {dial} that can be turned so as to span a specific sector of entangled steady states. 
Finally, we address an active protocol based on postselection, which projects the state of the qubits almost completely onto the symmetric state, thereby achieving nearly maximal entanglement. 

It is important to stress the differences between a bang-bang scheme~\cite{bangbang} and our own one. Although both scheme ultimately rely on the control of the interaction between the register and the environment, the two approaches are intrinsically different. Bang-bang schemes keep the state of interest unchanged throughout the evolution by effectively decoupling it from the environment. Our protocol however produces entanglement through the exploitation of purely dissipative dynamics. The protection of the entanglement from the influences of the external world is provided by the development of a {subradiant} behavior of the system due to the detuning modulation. 
Moreover, in a bang-bang protocol the timing is set by the fast switching rate of a control field, which is given by the inverse of the coupling strength between the system and the environment. Our scheme, on the other hand, is based on a weak-coupling regime between the register and the bus, which sets a slower time-scale than bang-bang schemes. This reflects not just a technical diversity, but is a manifestation of two almost complementary ways of designing protection from an environment. 

The remainder of the paper is organized as follows. In Section~\ref{system} we address the Bloch equations derived from the qubits' reduced master equation, in a weak-coupling regime and first Born-Markov approximation. This describes the dynamics of the qubits interacting with a leaky bus mode exposed to a bosonic multi-mode reservoir at thermal equilibrium with temperature $T$. The Bloch equations (as well as the master equation) fully account for a time-dependent detuning, which modulates the coupling between the qubits and the bus. Section~\ref{double detuning} explains in some detail how single-qubit addressing is not required in our scheme. This paves the way toward setting our proposal within the context of active (but simple) global-addressing scenarios. In Section~\ref{postselect} we describe the postselection protocol to improve the amount of entanglement set between the qubits by a nearly perfect projection of their joint state onto the symmetric state $\ket{s}$. Finally, Section~\ref{setup} describes a physical system to embody our proposed scheme. We show in some detail how a circuit-QED setup of two superconducting charge qubits incorporated in a microwave planar stripline resonator can be used. This offers practical advantages compared to the standard cavity-QED setup. The details regarding the derivation of the qubits' master equation are given in the Appendix.
\section{The system and detuning modulation protocol}
\label{system}

We consider two qubits, labelled $1$ and $2$, interacting with a one-sided single-mode cavity field described by the bosonic creation (annihilation) operator $\hat{a}^{\dag}$ ($\hat{a}$). Each qubit is characterized by a transition frequency $E^{0}$  and interacts with the cavity field (frequency $\omega_c$) via a dipole interaction with strength $g$. We assume a leaky cavity exposed to a multi-mode bosonic environment. The cavity mode is in a thermal state with temperature $T$ and average photon number $\bar{n}=(e^{\beta\omega_c}-1)^{-1}$, where $\beta^{-1}=K_{B}T$ and $K_B$ is the Boltzmann constant (we use units such that $\hbar=1$). The strength of the field mode-external bath coupling is given by the cavity decay-rate $\kappa$. Within the rotating wave approximation, we model each qubit-field interaction as $\hat{H}_{cj}=g(\hat{a}^{\dag}\hat{\sigma}^{-}_{j}+h.c.)$ ($j=1,2$), where $\hat{\sigma}^{-}_{j}=(\hat{\sigma}^{+}_{j})^{\dag}=\ket{0}_{j}\!\bra{1}$. The total energy of the system is 
\begin{figure}[t]
\psfig{figure=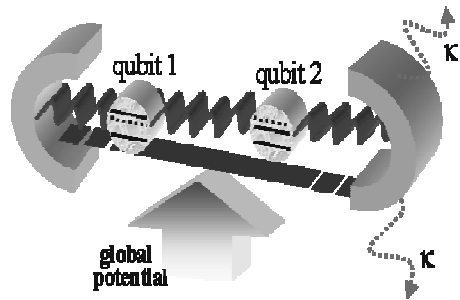,width=6.0cm,height=4.0cm}
\caption{(Color online) Two qubits interact with the same cavity mode (in contact with a thermal reservoir, decay rate $\kappa$) acting as a common bus. An inhomogeneous global potential, setting a time-dependent detuning is also shown.}
\label{scheme}
\end{figure}
\begin{equation}
\label{sistema}
\hat{H}_{sys}=\omega_c\,\hat{a}^\dag\hat{a}+\sum^{2}_{j=1}[\frac{E^{0}_{j}(t)}{2}\hat{\sigma}^{z}_{j}+\hat{H}_{cj}]
\end{equation}	
with $\hat{\sigma}^{z}_{j}$ the $z-$Pauli matrix of qubit $j$. We have considered an implicit time dependence of the single-qubit transition frequency. The detuning between the $j-$th qubit and the cavity field mode is indicated as $\Delta_{j}(t)=E^{0}_{j}(t)-\omega_{c}$. By introducing the transverse-mode energy decay rate $\gamma$ and assuming $\kappa\ll{\omega_{c}}$, the evolution of the qubits-bus system is described by the master equation Eq. (\ref{eq1}) given in the Appendix. Our system is sketched in Fig.~\ref{scheme}.

As we are interested in just the dynamics of the qubits, we adiabatically eliminate the bus mode and derive the corresponding Bloch equations. This is straightforwardly done by deriving a reduced master equation for the qubits only and projecting it onto the basis $\{\ket{\uparrow=11},\ket{s},\ket{a},\ket{\downarrow=00}\}_{12}$. We refer to the Appendix for the details of the adiabatic elimination. Here we concentrate on the form of the Bloch equations relevant to our work, where we assume the initial state $\ket{\uparrow}_{12}$ is prepared and, for the sake of simplicity, we consider the case of a detuning in amplitude smaller than the cavity decay rate. Using the notation $\varrho_{ij}=\mbox{}_{12}\expect{i}{\varrho}{j}_{12}$ ($i,j=\uparrow,s,a,\downarrow$) and $G^q_p(k\nbar)=q\gamma+\frac{g^2(k\nbar+p)}{\kappa}$ they read 
\begin{equation}
\label{bloch}
\begin{split}
&\partial_{t}\varrho_{\uparrow\uparrow}=-4G^1_1(\nbar)\varrho_{\uparrow\uparrow}
+2G^0_0(\nbar)(\varrho_{ss}+\varrho_{aa}),\\
&\partial_{t}\varrho_{ss}=-G^0_1(2\nbar)[\cos{\Delta(t)}\varrho_{ss}-i\sin{\Delta(t)}\varrho_{sa}+h.c.]\\
&+2G^1_1(\nbar)(\varrho_{\uparrow\uparrow}-\varrho_{ss})-2G^0_0(\nbar)(\varrho_{ss}-\varrho_{\downarrow\downarrow})\\
&+2\cos\Delta(t)[G^0_1(\nbar)\varrho_{\uparrow\uparrow}+G^0_0(\nbar)\varrho_{\downarrow\downarrow}],\\
&\partial_{t}\varrho_{sa}=-2G^1_1(2\nbar)\varrho_{sa}+i\sin{\Delta(t)}[2G^0_1(\nbar)\varrho_{\uparrow\uparrow}\\
&+2G^0_0(\nbar)\varrho_{\downarrow\downarrow}-G^0_1(2\nbar)(\varrho_{ss}+\varrho_{aa})].
\end{split}
\end{equation}
The equation for $\varrho_{aa}$ is given by that for $\varrho_{ss}$ with $\varrho_{ss}\rightarrow\varrho_{aa}$ and $\Delta(t)\rightarrow\pi-\Delta(t)$. The equation for $\varrho_{as}$ is the hermitian conjugate of the one for $\varrho_{sa}$ and the equation for $\varrho_{\downarrow\downarrow}$ is found from $\varrho_{\downarrow\downarrow}=1-\varrho_{\uparrow\uparrow}-\varrho_{ss}-\varrho_{aa}$. We have taken $\Delta_1(t)>0$, with $\Delta_{2}(t)=0$ and $\Delta_{1}(t)=\Delta(t)$ for ease of notation. In Section~\ref{double detuning} we consider the generalization of this situation to two-qubit detuning and in the Apppendix we provide the form of the reduced master equation. Moreover, we stress that the absence of terms like $\varrho_{\uparrow{a}},\varrho_{\uparrow{s}}$ (and analogous) is due to the specific choice of the initial state. In particular, if any coherence is initially present in the qubit state, the set of Eqs.~(\ref{bloch}) must be complemented by a second closed system of Bloch equations which can easily be derived. 

From Eqs.~(\ref{bloch}) the initial state $\ket{\uparrow}_{12}$
evolves into
\begin{equation}
\label{density}
\varrho=\sum_{j=\uparrow,a,s,\downarrow}\varrho_{jj}\ket{j}_{12}\!\bra{j}+(\varrho_{as}\ket{a}_{12}\!\bra{s}+h.c.).
\end{equation}

\begin{figure}[t]
\psfig{figure=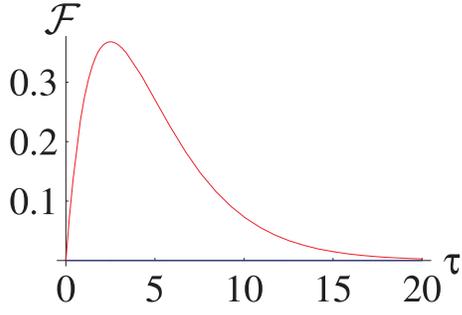,width=6.cm,height=4.0cm}
\caption{(Color online) Fidelity of the two-qubit state $\varrho$ for the full-resonant case with the symmetric state $\ket{s}_{12}$ (solid line) and the antisymmetric one $\ket{a}_{12}$ (identical to zero) against the dimensionless interaction time $\tau=g^2t/\kappa$. 
}
\label{fedelta}
\end{figure}
In order to illustrate the basic features of our proposal, we consider $\gamma=\bar{n}=0$ for the moment. These parameters will be re-introduced later on. As a measure of the entanglement in the bipartite mixed state of qubits $1$ and $2$, we use the concurrence~\cite{wootters} which, for a bipartite state, 
can be calculated as ${\cal C}(\varrho)=\max[0,\sqrt{\alpha_{1}}-\sum_i\sqrt{\alpha_i}]$ with $\bar{\varrho}={\varrho(\otimes^{2}_{j=1}\sigma^{j}_{y})\varrho^{*}(\otimes^{2}_{j=1}\sigma^{j}_{y}})$ ($\varrho^{*}$ is the complex conjugate of $\varrho$) and $\alpha_1\ge\alpha_i\,(i=2,3,4)$ are the eigenvalues of $\bar{\varrho}$. For a qubit state as Eq.~(\ref{density}), we have
\begin{equation}
\label{concurrence}
{\cal C}(\varrho)=\max[0,2(\vert\varrho_{10,01}\vert-\sqrt{\varrho_{\uparrow\uparrow}\varrho_{\downarrow\downarrow}})]
\end{equation}
with $2\varrho_{10,01}=\varrho_{ss}+\varrho_{sa}-\varrho_{as}-\varrho_{aa}$. In order to gain as much information as we can about the behavior of ${\cal C}(\varrho)$, we relax the $\max$ condition. Thus, in the following plots, entanglement is present only when ${\cal C}(\rho)>0$. Moreover, we address a physically interesting situation by considering the initial state $\ket{\uparrow}_{12}$ which, as our system is formally equivalent to a Dicke model~\cite{dicke}, decays toward the ground state $\ket{\downarrow}_{12}$ on a time-scale which is faster than the single-qubit relaxation time. Thus, for the steady state of the system, no entanglement is expected between $1$ and $2$. As we stress later, this choice for an initial state is dictated by the global addressing context of this work. The state $\ket{\uparrow}_{12}$ can be prepared via a global potential and without local control~\cite{commento}. 


As time passes, the superradiant state is rotated toward a mixed state. While still exhibiting no quantum correlations (it never violates the Peres-Horodecki separability criterion~\cite{PH}), the state nevertheless has a good projection onto $\ket{s}_{12}$ and no contribution from the antisymmetric state $\ket{a}_{12}$. This is shown in Fig.~\ref{fedelta} where the fidelities ${\cal F}_{j}=\mbox{}_{12}\expect{j}{\varrho}{j}_{12}=\varrho_{jj}$ ($j=s,a$) are plotted against the dimensionless interaction time $\tau=g^2t/\kappa$.
At $\tau\simeq{2.5}$, $\varrho_{ss}\gtrsim{0.37}$ (with $\varrho_{aa}=0$), though the state is still mixed and separable due to the non-zero value of the fully polarized states $\ket{\uparrow}_{12}$ and $\ket{\downarrow}_{12}$. This suggests a minimization of the influence of these components on~$\varrho$. The idea behind our proposal is to exploit this fact and change the dynamical evolution of the qubits, by introducing a detuning $\Delta$, so as to induce an evolution with an initial state which is no more $\ket{\uparrow}_{12}$ but $\varrho(\tau=2.5)$. 
\begin{figure}[b]
\psfig{figure=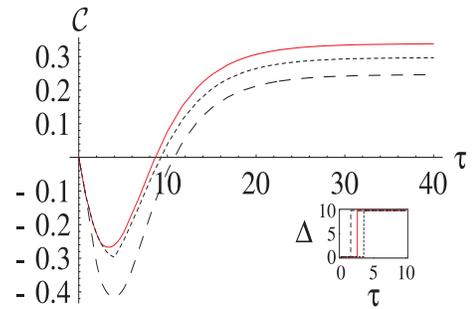,width=6.cm,height=4.0cm}
\caption{(Color online) ${\cal C}(\rho)$ against $\tau$ for the initial state $\ket{\uparrow}_{12}$ and $\Delta(\tau)=10\Theta(\tau-\tau_0)$. We have considered $\tau_0=1.5, 2.5, 3.5$ (dashed, solid and dotted line respectively). The inset shows the behavior of the detuning functions.}
\label{condetu}
\end{figure}
Thus, at $\tau=2.5$ an external potential is switched on and changes the transition frequency $E^{0}_1(t)$. The dynamics of the qubits are therefore described by Eqs.~(\ref{bloch}) (no component outside the subspace encompassed by Eqs.~(\ref{bloch}) is present in the new initial state) and the concurrence is plotted against $\tau$ in Fig.~\ref{condetu} (solid line), where $\Delta=10g^2/\kappa$ with $g=0.3\kappa$, so that the adiabatic condition is fully respected. We find ${\cal C}>0.3$, stable at the steady state of the qubits. This interesting result can be compared with~\cite{myungpeter} which does not achieve a stable entanglement. The plot results from a transition between the entanglement function of the full resonant case before $\tau=2.5$ and that of the single-qubit detuning described above for $\tau>2.5$. This situation is equivalent to taking $\Delta(\tau)=10\Theta(\tau-2.5)$ (in units of $g^2/\kappa$), where $\Theta(\tau-\tau_0)$ is the Heaviside function. The steady-state value of the entanglement turns out to dependent weakly on the amplitude of $\Delta(\tau)$ but strongly on $\tau_0$. For instance, in Fig.~\ref{condetu} we show the results of small deviations from the case considered above by plotting the concurrence relative to $\tau_0=1.5$ (dashed line) and $\tau_0=3.5$ (dotted line), which give rise to smaller steady-state entanglement. This results from a smaller $\varrho_{ss}(\tau_0)$ component in the two-qubit state and a disadvantageous competition between $\ket{s}_{12}$ and $\ket{a}_{12}$, which lowers the entanglement. This is strikingly exemplified by increasing $\tau_0$ by one order of magnitude. In this case, the switching on of the detuning occurs in correspondence to a state of the two qubits which is mainly decayed to $\ket{\downarrow}_{12}$ ($\varrho_{\downarrow\downarrow}(\tau_0=25)=0.999501$).

It is worth stressing that, even though our choice for $g/\kappa$ may seem to put the above example at the boundary of the applicability of the adiabatic elimination, we have checked that for $g/\kappa\sim{0.1}$, no significant change occurs in the entanglement generation process~\footnote{The main implication of choosing a smaller $g/\kappa$ ratio is the shift toward larger values of the instant at which the concurrence starts to be positive.}. In principle, the true evolution of the system, obtained by numerically solving the complete master equation (without adiabatic elimination) should be compared to the situation here at hand. However, this is in general a very hard task, which goes beyond the scopes of the present work. 
Nevertheless, by checking the effects of values for $g/\kappa$ which are largely within the validity of the adiabatic elimination, we can be sure of the validity of the above approach. The system is flexible enough to tolerate a less strict ratio of the different time-scales involved in the problem.

\begin{figure}[t]
\psfig{figure=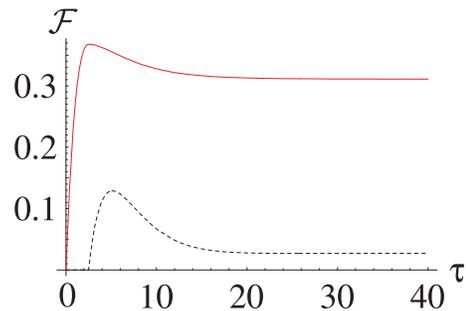,width=6.0cm,height=4.0cm}
\caption{(Color online) Fidelities ${\cal F}_{s,a}$ for a detuning modulation strategy based on a Heaviside function with $\tau_0=2.5$. The solid line is for ${\cal F}_s$, the dotted one for ${\cal F}_a$. At $\tau=\tau_0$, an antisymmetric state component is developed. For large $\tau$, both the antisymmetric and symmetric state fidelities are stabilized.}
\label{fedelta2}
\end{figure}

The appearance of steady-state entanglement can be shown clearly by the behavior of the density matrix elements.
A careful analysis reveals that the qubit entanglement is due to the presence of $\ket{s}_{12}$ and $\ket{a}_{12}$. The calculation of the fidelities ${\cal F}_{s,a}$ in presence of the detuning modulation, reveals that as soon as the detuning is switched on, an $\ket{a}_{12}$ component is developed, which after a transient period, stabilizes to a steady state value. This stabilization holds also for the $\ket{s}_{12}$ component, with $\varrho_{ss}\gg\varrho_{aa}$. These behaviors can be seen in Fig.~\ref{fedelta2} for the Heaviside function with $\tau_0=2.5$. However $\varrho_{\downarrow\downarrow}$ never vanishes, thus affecting the entanglement. 
This clarifies the mechanism behind the entanglement generation and protection. Without a detuning modulation, the system would never develop any {subradiant} behavior ({\it i.e.} the overlap with $\ket{a}_{12}$ will always be zero). This is not true for a modulated situation, where the dynamical conditions are changed. Once the system has decayed into an incoherent superposition of $\ket{s}_{12}$, $\ket{\uparrow}_{12}$ and $\ket{\downarrow}_{12}$ (for $\tau<\tau_0$), one qubit acquires a relative phase with respect to the other one due to the detuning, which results in the development of a subradiant component. The steady state entanglement is the result of a competition between $\ket{\downarrow}_{12}$, the antisymmetric and the symmetric component. 

\begin{figure}[b]
\psfig{figure=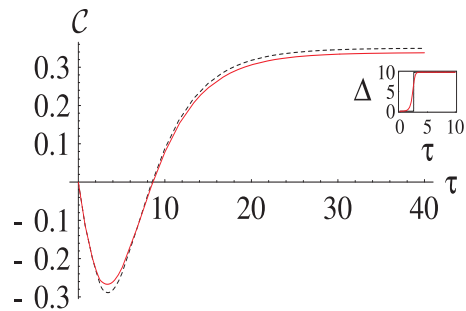,width=6.0cm,height=4.0cm}
\caption{(Color online) Comparison between the concurrence obtained by using a Heaviside function $10\Theta(\tau-2.5)$ (dashed line) and the smooth detuning function ${10}[{{1+e^{6(2.5-\tau)}}}]^{-\frac{1}{2}}$ (solid line). The inset shows the time behavior of the detuning functions.}
\label{confrontodetu}
\end{figure}

The assumption of a Heaviside function regulating $\Delta(t)$ is not critical. In order to relax this assumption, we have checked the results corresponding to a smooth raising edge of the detunings given by the function ${\cal A}[{{1+e^{2b(\tau_0-\tau)}}}]^{-\frac{1}{2}}$ with ${\cal A}$ an amplitude.
For proper choices of ${\cal A}$ and $b$, this is a slowly rising function (with respect to the time-scale set by $\kappa$, see the Appendix) producing a concurrence which differs from the result obtained for a Heaviside function by less than $1\%$, as shown in Fig.~\ref{confrontodetu} for $b=3,\tau_0=2.5$ and ${\cal A}=10$.
Thus, for definiteness and in order to simplify the calculations, we assume a Heaviside profile of the detuning rising edge~\footnote{The considerations made about the ratio $g/\kappa$ hold also in relation to the smooth rising-edge considered here. A smaller ratio does not change the validity of the conclusions drawn so far.}.

It should now be clear that a detuning function with a single rising edge represents the best choice. This can be confirmed by considering the value of ${\cal C}$ as a function of the time width of a single square pulse. As already stressed, the turning on of the detuning corresponds to  the maximization of the symmetric state component and the introduction of an $\ket{a}_{12}$ component in $\varrho$. If the detuning is switched off after a time $\delta\tau$, the symmetric component quickly goes to zero (together with any correlation betwen $\ket{s}$ and $\ket{a}$) while the subradiant part is preserved. This achieves a non-zero stationary entanglement which nevertheless is reduced with respect to the case of a Heaviside function. Indeed, in the above conditions, at the generic time $\tau_e$, the steady-state entanglement quantitatively corresponds to the fraction of the antisymmetric state being present in $\varrho$, as can be immediately seen by considering a state like $\varrho=\sum_{j=\uparrow,s,a,\downarrow}A_j\ket{j}_{12}\!\bra{j}$.
For $\tau_e-\tau_0\gg\delta\tau$, the population of $\ket{\uparrow}_{12}$ is zero and so is the symmetric state fraction. That is $A_{\uparrow,s}=0$ so that $C(\varrho)=A_a$. The entanglement is stable, even though small, due to the sole presence of the subradiant component, developed at the switching-on of the pulse. As soon as $\delta\tau$ becomes larger than $\tau_e$, making the symmetric state fraction (and its correlations with $\ket{a}_{12}$) non-negligible, the entanglement is not only stable but also reaches the asymptotic value corresponding to Fig.~\ref{condetu}, as can easily be seen. As $\tau_e$ is increased, this behavior holds for a larger $\delta\tau$, demonstating that a large steady-state entanglement is achieved only for a step-like function (as mentioned before, the raising edge functional behavior is irrelevant).

\begin{figure}[b]
\psfig{figure=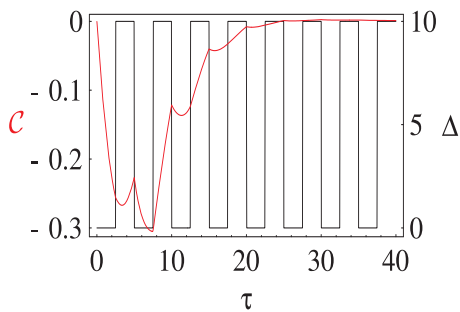,width=6.0cm,height=4.0cm}
\caption{(Color online) Concurrence (red curve, left vertical scale) superimposed on the detuning function $\Delta_{sw}(\tau)$ (black curve, right scale) against $\tau$ for the qubits initially prepared in the superradiant state $\ket{\uparrow}_{12}$. In this plot we have considered $N=16$. No entanglement is ever present between the qubits.}
\label{ondaquadra}
\end{figure}

A further case can be considered, namely a periodic modulation. However, this modulation implies the switching on/off of the detuning at instants of time that correspond to smaller fidelities of the state $\varrho$ with the symmetric state. For instance, in Fig.~\ref{ondaquadra} we consider the case of a square wave 
$\Delta_{sw}(\tau)=10\sum^{N}_{n=1}(-1)^{n+1}\Theta(\tau-2.5 n)$,
which produces $N/2$ square pulses of amplitude $10$ (units of $g^2/\kappa$). 
No entanglement results from this detuning modulation strategy, as the switching of the detuning introduces a jagged-drop in the fidelity compared to the situation in Fig~\ref{fedelta2}. While the on part of the square wave always corresponds to a slow {increase} of ${\cal C}$, the off part results in a larger decrease, resulting in an overall {\it pull down} of entanglement. 

\begin{figure}[b]
\psfig{figure=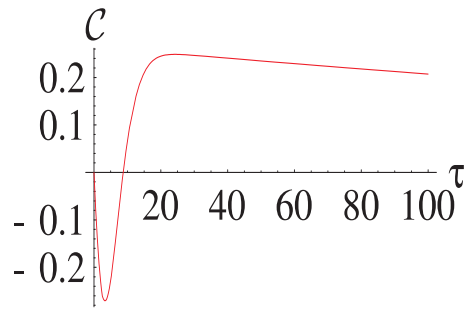,width=6.0cm,height=4.0cm}
\caption{(Color online) Concurrence against $\tau$ for a modulated detuning with $g=0.3\kappa,\,\gamma=10^{-3}\kappa$ and $\bar{n}=0.06$. The time axis has been extended in order to show that a concurrence close to $0.2$ is present for $\tau$ up to $100$. The slow entanglement decay is due to the presence of a non-zero $\gamma$, while the smaller steady-state value is largely due to the thermal nature of the bath ($\bar{n}\neq{0}$).}
\label{condecrease}
\end{figure}

So far, only the case with no spontaneous emission and a zero-temperature bath has been considered. In order to include the effects of $\gamma,\nbar\neq{0}$, we need to solve Eqs.~(\ref{bloch}) for $\gamma,\bar{n}\neq{0}$ and relate them more closely to a physical setup that will implement our protocol. Although more detail will be given later, here we mention that the situation considered is such that $\gamma\ll{g^2/\kappa}$ and $\bar{n}\ll{1}$, which are realistic conditions in several physical systems such as circuit-quantum electrodynamics of superconducting charge qubits integrated in microwave cavities~\cite{ioJJ,schoelkopf}. We will postpone discussion of the order of magnitude of these physical parameters until Section~\ref{setup}. To fix the ideas and to be as close as possible to physical reality, we consider $\bar{n}=0.06$ and $\gamma=10^{-3}\kappa$. Moreover, in tackling this analysis, we find it convenient to refer to the computational basis $\{\ket{\uparrow},\ket{10},\ket{01},\ket{\downarrow}\}_{12}$. The essential features of the previously considered case still hold, with the ${\cal F}_s$ function also maximized at $\tau=2.5$. The main effect of the non-zero thermal photon number is a reduction in the steady-state entanglement value. This is due to $\bar{n}\neq{0}$, which reduces the coherence $\varrho_{10,01}$, thus preventing the state from mimicking the symmetric state. On the other hand, $\gamma\neq{0}$ introduces a second time-scale in the system, which results in a slow decay of both the populations of states $\ket{10}_{12},\,\ket{01}_{12}$ and of the coherence $\varrho_{10,01}$. This accounts for an overall decay of ${\cal C}$, as shown in Fig.~\ref{condecrease}.
Despite the decrease found for non-zero $\gamma$, the entanglement still remains as large as $0.2$ for interaction times up to $\tau=100$. This corresponds to an effective entanglement protection from both thermal behavior of the bosonic bath and the qubits' spontaneous decay.

\section{Double detuning and global addressing}
\label{double detuning}

In this Section we address the question of whether the assumption of a single-qubit detuning modulation is critical to the proposed protocol. The answer to this provides an effective justification and an {\it a posteriori} motivation for the analysis conducted so far. In order to do this, we re-consider the physical system described in Section~\ref{system} (see the Appendix) with the inclusion of the second qubit detuning $\Delta_2(t)$. This modifies the dynamics of the system in such a way that the Bloch equations (\ref{bloch}) remain unaltered with the replacement ${\Delta}(t)\rightarrow\tilde{\Delta}(t)=\Delta_{1}(t)-\Delta_{2}(t)$. Thus, considering both the qubits as detuned in time is equivalent to just considering the energy of qubit $1$ being modulated with an effective time-dependent detuning $\tilde\Delta(t)$. That is, the analysis conducted so far is perfectly general and there is no limitation in considering a single-qubit modulation as this case encompasses rigorously the most general situation of double detuning. Obviously, the $\Delta_{j}(t)$'s cannot be chosen arbitrarily, in general. 

This result has two main implications. The first is that in order for the protocol we have described to be effective, we must consider detuning functions which are opposite in sign ({\it i.e.} one detuning has to be positive, the other negative). The second point is pragmatically relevant as we can now put our scheme within the context of global addressing protocols~\cite{sougato}. Indeed, it should be clear after the above discussion that the realization of the detuning-modulation protocol simply requires the appropriate setting of a potential which addresses both the qubits in the correct way (increasing the energy spacing of qubit $1$ with respect to the resonant value $\omega_c$ and reducing the spacing of qubit $2$). No single-qubit addressability is needed, which considerably reduces the experimental efforts for the implementation of the scheme. It is not necessary to require a strongly focused potential applied to just one of the two qubits and having no effect on the dynamics of the other one. In order to fix this idea, one can consider a global magnetic field inducing a {Zeeman-like} effect on the qubits' energy levels, the shifting being different from qubit to qubit because of a gradient in the magnetic potential (see Fig.~\ref{scheme}). In Section~\ref{setup} we address the physical mechanism responsible for such a shift by considering a specific experimental setup that can be used in order to implement our proposal. 
\section{Entanglement improvement by postselection}
\label{postselect}

\begin{figure}[t]
\psfig{figure=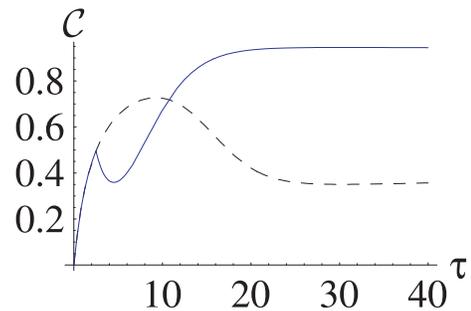,width=6.0cm,height=4.0cm}
\caption{(Color online) Comparison between the entanglement in the postselected state with and without the detuning-modulation protocol (solid and dashed curves respectively). A Heaviside modulation is considered in the solid line case.}
\label{entpost}
\end{figure}

As the entanglement is set in a (quasi) steady-state, the required degree of control over the system is reduced. With the exception of the choice for the optimal value of $\tau_0$ at which to switch on the detuning, no fine time control is necessary in order to properly drive the dynamics of the system. Nevertheless, it might be desirable in many situations to raise ${\cal C}$ up to a maximal entanglement of one ebit. We have seen an intrinsic limitation in the amount of establishable entanglement due to the unavoidable presence of the spurious population of $\ket{\downarrow}_{12}$. 
A procedure which allows us to cut away the unwanted contribution from $\ket{\downarrow}_{12}$ is represented by the postselection of the two-qubit state after some detection event. Explicitly, consider the fading influence of the $\varrho_{\uparrow\uparrow}$ component. In the specific case here at hand, each time the state of the two qubits is not found to be $\ket{\downarrow}_{12}$, the overlap with $\ket{s}_{12}$ increases, improving the entanglement between the qubits. Thus, by using the positive-operator-valued-measure (POVM) $\{\hat{\Pi}_{0}=\ket{\downarrow}_{12}\!\bra{\downarrow},\hat{\Pi}_{1}=\openone-\hat{\Pi}_{0}\}$ with $\openone$ the identity operator, we can postselect the state resulting from the qubits not being found in the global ground state. This changes $\varrho$ into 
$\varrho_{p}={\cal N}\hat{\Pi}_{1}\varrho\hat{\Pi}_{1}={\cal N}(\varrho-\varrho_{\downarrow\downarrow}\ket{\downarrow}_{12}\!\bra{\downarrow})$ with ${\cal N}$ a normalization factor. As $\varrho_{\uparrow\uparrow}\rightarrow{0}$, this effectively results in a projection of the two-qubit state onto the subspace spanned by $\ket{s,a}_{12}$ with asymptotically $\mbox{}_{12}\expect{s}{\varrho_p}{s}_{12}>{0.9}$. After the analysis in Section II, we know that a large fraction of $\ket{s}_{12}$ implies a large degree of entanglement. This is witnessed by the entanglement properties of the resulting state, which is represented in Fig.~\ref{entpost}. The plot represents the amount of entanglement in the postselected state when the measurement is performed at the instant $\tau$ in the evolution of the two-qubit state. Both the detuning-modulated (solid line) and the full-resonant case (dashed line) are shown. In both the cases there is an improvement in the amount of stationary entanglement.
While the second case has concurrence stabilized around $0.4$, an entanglement larger than $0.9$ is obtained in the modulated case. However, a full ebit is not possible as the spontaneous emission and the thermal effects of the bath spoil the correlations between the qubits. Indeed, we have checked that almost a complete ebit is achievable if $\gamma=\bar{n}=0$ is considered, where the detuning-modulated condition is better than the unmodulated case, as the concurrence value always remains above the entanglement curve of the unmodulated case. 


\section{Physical setup}
\label{setup}

The physical system we suggest to use in order to implement our proposal is given by two superconducting charge qubits, embodied by Superconducting-Quantum-Inferference-Devices (SQUIDs)~\cite{schon}, nano-lithographically implanted in a quasi-unidimensional microwave stripline resonator~\cite{zmuidinas}. This system offers advantages in many respects. First, the qubits are stationary within the cavity, so that the requirement for a fine tuning of the transit-time through the cavity (typical of microwave cavity-QED implementations) is no more an issue. Second, the coupling between the qubits in the register and the cavity bus can easily be arranged so as to satisfy the weak coupling regime required by our proposal. Finally, the manipulability of charge qubits embodied by SQUIDs allows for a detuning modulation in the global addressing fashion depicted in this paper. 

In detail, we assume the charging regime and the low-temperature limit~\cite{schon} and set each SQUID to work at the charge degeneracy point, where the qubits are encoded in equally-weighted superpositions of states, having zero and one excess Cooper-pair on the SQUID island, namely $\ket{\pm}_j=(1/\sqrt{2})(\ket{0}\pm\ket{2e})_j$ ($2e$ being the charge of a Cooper pair). The degeneracy point is set by biasing each SQUID with a dc electric field connected to the superconducting devices via the ground plate of the resonator. The free Hamiltonian of a single SQUID is thus given by $(1/2)E^{0}_{j}(\phi)\hat{\sigma}^{z}_{j}$, with $E^{0}_{j}(\phi)$ the Josephson energy (tuned via an external magnetic flux $\phi$ piercing the SQUIDs). By modulating the magnetic flux, we can change the energy separation between the qubit levels, thus setting the detunings with respect to the cavity mode frequency. A gradient can be incorporated into the external magnetic flux so as to realize a configuration of equal and opposite detunings for the qubits in our register. 

The microstrip resonator can be modelled as a distributed $LC$ oscillator, where $C$ is the capacitance between the plates of the stripline and $L$ is the overall inductance of the device (depending on the length of the resonator, typically in the range of $1$ cm). In this setup, $\gamma/\kappa\simeq{10^{-3}}$ and a cavity quality factor of $\sim100$ are conservative assumptions.
At $\omega_{c}/2\pi=6$ GHz and $T\simeq170$ mK we have $\bar{n}\simeq{0}.06$. The coupling between qubits and cavity mode is capacitive and mediated by the electric part of the cavity field~\cite{ioJJ,schoelkopf}. In a second quantization picture, the interaction Hamiltonian can be cast in the form of a Jaynes-Cummings model so that $\hat{H}_{sys}$ in Eq.~(\ref{sistema}) can naturally be embodied by the present setup. A Liouvillian description of SQUID-cavity open systems has been proven to be rigorous up to temperatures well above those assumed in this work. Indeed, for two qubits in a stripline resonator, the optical master equation~(\ref{eq1}) can be derived from the Bloch-Redfield formalism, when the secular approximation is relaxed and a large number of elements of the Redfield tensor are considered~\cite{rau}.

Two SQUID qubits (size $\sim\mu$m) can easily be accommodated in the cavity far enough away from each other to achieve negligible cross-talk (in principle due to direct capacitive and inductive coupling). Lithographic techniques allow us to control, within a few percent, the geometric characteristics and the resulting parameters of the device. The two qubits can therefore be manipulated both simultaneously or independently with two separate coils. Due to charged impurities in the vicinity of the devices, separate calibration at the degeneracy points would be required for each qubit. This may be achieved with several adjustments to the design of the setup. For instance, by splitting the ground plate and attaching a gate to each part~\cite{ioJJ}. 

Let us now turn briefly to the description of possible ways of implementing the conditional detection scheme described in Section~\ref{postselect}. In principle, a measurement of the qubits' state can be performed by setting a large qubit-cavity field detuning, attaching a detector at the output capacitive gap of the stripline resonator and measuring the shifts induced in the resonance spectrum of a probe beam sent into the cavity through the input capacitive gap. The dispersive nature of the qubit-cavity coupling, which changes the refractive index of the cavity field mode, determines qubit-state dependent shifts in the resonance peak of the probe beam. This allows for the non-demolition detection of the qubit state, following the strategy depicted in~\cite{schoelkopf}. However, in order for these shifts to be detectable, the change of the refractive index has to be larger than the cavity linewidth, a condition which is hard to match if the bad cavity regime is invoked. However, a second strategy is possible, which is more suitable for conditions of large detunings between the cavity and register and a large cavity decay rate. This involves driving a cavity field mode with a coherent state $\ket{\alpha}$ ($\alpha\in{\mathbb C}$). In the situation of a large qubit-cavity detuning, the dispersive dynamics the system undergoes is such that the globally unexcited state $\ket{\downarrow}_{12}$ becomes correlated with the field state $\ket{\alpha{e}^{i\theta}}$~\cite{ionjp,schoelkopf}. That is, in phase space the coherent state acquires an additional phase dependent in general on the ratio $2g^2/\Delta$. On the other hand, the symmetric and antisymmetric component of the density matrix leave the coherent state unchanged~\cite{ionjp}. A homodyne measurement of the cavity field provides a distinction between the states of the register and therefore the implementation of the POVM we have described. 

As an additional remark, we stress that in this setup, at the charge degeneracy point, decoherence due to low-frequency modes vanishes at the first order. This allows the minimization of the effects of noise sources represented by switching charged impurities in the proximity of the SQUIDs' islands, which constitute a system of bistable fluctuators giving rise to $1/f$ noise~\cite{elisabetta}. Finally, it is worth stressing that due to the qubit-resonator interaction, the energy levels of our qubits are much less sensitive to these charge fluctuations than isolated qubits at the optimal working point~\cite{ioJJ}. This allows us to neglect any resulting dephasing effects. 

Finally, it is worth mentioning that the example considered in this Section is just one of the possible physical setups where our proposal could be implemented. Indeed, the formalism we have used in order to describe the main features of our protocol is general enough to be adapted to various situations. For instance, the case of two trapped ions, in the Lamb-Dicke regime and placed inside an optical cavity can be taken in consideration~\cite{walther}. The extension of our analysis to the case of multi-level systems composing the register, on the other hand, will pave the way to the use of two closely-spaced ensembles of cold two-level atoms (confined in vapor cells or magneto-optical traps). The free-space interaction of a laser with the ensembles, each treated as an effective $N/2$-spin (where $N$ is the number of atoms in each ensemble) and within the rotating wave approximation, provides an interaction Hamiltonian which is the generalization of our model to $N+1$ systems~\cite{polzik}.

\acknowledgments

We thank  the Leverhulme Trust (ECF/40157), the UK EPSRC, KRF (2003-070-C00024) and DEL for financial support. MP thanks C. Brukner and J. Kofler for useul discussions.

\appendix
\section*{Appendix: Master equation}
\label{appendice}
\subsection*{Single-qubit detuning}

In this Appendix we briefly describe the steps taken in order to derive the Bloch equations~(\ref{bloch}). The last part of the Appendix is dedicated to the case of a double detuning, as studied in Section~\ref{double detuning}. In an open-system perspective, the dynamics of the cavity field-qubit system is described by the master equation
\begin{equation}
\label{eq1}
\partial_{t}\rho_{c12}=-i[\hat{H}_{sys},\rho_{c12}]+\hat{{\cal L}}^{\kappa}_c[\rho_{c12}]+\sum^{2}_{j=1}\hat{{\cal L}}^{\gamma}_j[\rho_{c12}],
\end{equation}
where $\hat{H}_{sys}$ is defined by Eq.~(\ref{sistema}).
Here, we have introduced the Liouvillian superoperators describing the cavity decay $\hat{{\cal L}}^{\kappa}_c[\rho_{c12}]=\kappa(\bar{n}+1)[2\hat{a}\rho_{c12}\hat{a}^{\dag}-\{\hat{a}^{\dag}\hat{a},\rho_{c12}\}]+\kappa\bar{n}[2\hat{a}^{\dag}\rho_{c12}\hat{a}-\{\hat{a}\hat{a}^{\dag},\rho_{c12}\}]$ and the single-qubit spontaneous emission $\hat{{\cal L}}^{\gamma}_j[\rho_{c12}]$, which has the same structure as $\hat{{\cal L}}^{\kappa}_c[\rho_{c12}]$ for $\bar{n}=0,\,\kappa\to\gamma$ and $\hat{a}\to\hat{\sigma}^{-}_{j}$. 

As we are interested in the qubits' evolution, we now adiabatically eliminate the degrees of freedom of the field mode. This can be done using a procedure valid in the weak-coupling regime $g\ll{\kappa}$. In this case, the dynamics of the mode interacting with the bath is much faster than its interaction with the qubits. The qubits see the cavity mode
in a steady state $\rho_{ss}$ not affected by the qubit-mode dynamics and determined by the statistical properties of the environmental bath. In the case at hand, $\rho_{ss}$ is a thermal state with average photon number $\bar{n}$. By using, standard techniques of quantum optics based on second-order perturbation theory and the first Born-Markov approximation~\cite{wallsmilburn} (see also ref.~\cite{tanasficek}), the qubit master equation in the interaction picture reads
\begin{equation}
\label{uno}
\partial_{t}\varrho=\sum^{2}_{j=1}(\hat{\mathcal L}^{g^2/\tilde\kappa_j}_{j}[\varrho]+\hat{{\cal L}}^{\gamma}_j[\varrho])+\hat{\mathcal L}^{g^2/\kappa}_{12}[\varrho]
\end{equation}
with 
\begin{equation}
\label{singolo}
\begin{split}
\hat{\mathcal L}^{{g^2}/{\tilde\kappa_j}}_{j}[\varrho]&=\frac{g^2}{\tilde\kappa_j}(\bar{n}+1)\left[2\hat\sigma^{-}_{j}
\varrho\hat\sigma^{+}_{j}-\{\hat\sigma^{+}_{j}\hat\sigma^{-}_{j},\varrho\}\right]\\
&+\frac{g^2}{\tilde\kappa_j}\bar{n}\left[2\hat\sigma^{+}_{j}\varrho\hat\sigma^{-}_{j}-\{\hat\sigma^{-}_{j}\hat\sigma^{+}_{j},\varrho\}\right],
\end{split}
\end{equation}
\begin{equation}
\label{doppio}
\begin{split}
\hat{\mathcal L}^{{g^2}/{\kappa}}_{12}[\varrho]&\simeq\frac{g^2(t)}{\kappa}(\bar{n}+1)[2\hat\sigma^{}_{1-}
\varrho\hat\sigma^{+}_{2}-\{\hat\sigma^{+}_{2}\hat\sigma^{-}_{1},\varrho\}]\\
&+\frac{g^2(t)}{\kappa}\bar{n}[2\hat\sigma^{+}_{2}
\varrho\hat\sigma^{-}_{1}-\{\hat\sigma^{-}_{1}\hat\sigma^{+}_{2},\varrho\}]+h.c.,
\end{split}
\end{equation}
and $g(t)=ge^{-\frac{i}{2}\Delta(t)}$. Here, we have considered the case of $E^{0}_{1}(t)\gg{\omega}_{c}$ with $E^{0}_{2}(t)\equiv{E}^{0}_{2}=\omega_c$ (so that $\Delta_{2}(t)=0$ always) and $\Delta_{1}(t)=\Delta(t)$ for ease of notation. Eq.~(\ref{singolo}) is the Liouvillian describing the single-qubit decay rate induced by the coupling to the external thermal bath mediated by the bus. In general, the cavity-induced decay rate of atom $j$ depends on the corresponding detuning. 
Within the range of validity our approximations, however, the bare decay rates can be used. 
It is worth stressing that as the adiabatic elimination procedure does not affect the qubits degrees of freedom, the form of $\hat{{\cal L}}^{\gamma}_j[\varrho]$ is left unchanged. However, the adiabatic elimination of the cavity gives rise to an effective qubit-qubit interaction term.
The structures of Eqs.~(\ref{singolo}) and (\ref{doppio}) result from the assumption that the detuning $\Delta(t)$ is modulated within a time-scale slower than the one set by the cavity decay rate $\kappa$. 
\subsection*{Two-qubit detuning}
The introduction of the second-qubit detuning proceeds as described in Section~\ref{double detuning} and by re-considering Eq.~(\ref{uno}). By revising the adiabatic elimination procedure, the form of the single-qubit decaying terms in Eq.~(\ref{singolo}) remain unaltered, while the inter-qubit correlation term is changed into
$\hat{\mathcal L}^{{\tilde{g}^{2}}/{\kappa}}_{12}[\varrho]$
with $\tilde{g}^{}(t)=ge^{-\frac{i}{2}\tilde{\Delta}(t)}$, $\tilde\Delta(t)=\Delta_{1}(t)-\Delta_{2}(t)$ and $\Delta_{j}(t)$ the detuning of qubit $j$. Evidently, the form of the master equation is invariant after the introduction of the two-qubit detuning and the dynamics depend only on the relative detuning between the qubits. This is due to the structure of Eq.~(\ref{eq1}) in the adiabatically eliminated form. In $\hat{\mathcal L}^{{\tilde{g}^2}/{\kappa}}_{12}[\varrho]$, the presence of a lowering ladder operator ({\it i.e.} $\hat\sigma^{-}_{j}$) is accompanied by the rising ladder operator of the other qubit. As a lowering operator is associated with the time-dependent term $e^{-i\Delta_{1}(t)}$, the accompanying rising operator will introduce $e^{i\Delta_{2}(t)}$ so that they always combine to give the relative detuning $\tilde{\Delta}(t)$.


\begin{thebibliography} {99}
\bibitem{commento} We have checked that initial states $\ket{j,1-j}_{12}$ ($j=0,1$) do not change the quantitative and qualitative picture.
\end{thebibliography}
\end{document}